\documentclass[twocolumn,trackchanges]{aastex62}
\usepackage{amsmath}
\hypersetup{
	colorlinks	= true,
	linkcolor	= red,
	urlcolor	= cyan,
	citecolor	= blue
}

\usepackage{amssymb}

\usepackage{lipsum}
\usepackage{multirow}

\graphicspath{{./images/}}

\newcommand{\sister}{\mbox{\textsc{SISTER}}}
\newcommand{\exorelr}{\mbox{\textsc{ExoReL$^\Re$}}}
\newcommand{\exorel}{\mbox{\textsc{ExoReL}}}
\newcommand{\umab}{\mbox{47 UMa b}}

\turnoffediting

\received{July 7, 2020}
\accepted{September 8, 2020}

\submitjournal{AJ}

\shortauthors{Damiano et al.}

\usepackage{fancyhdr}
\begin{document}
	
	\title{Multi-orbital-phase and multi-band characterization of exoplanetary atmospheres with reflected light spectra}
	
	\correspondingauthor{Mario Damiano}
	\email{mario.damiano@jpl.nasa.gov}
	
	\author[0000-0002-1830-8260]{Mario Damiano}
	\affiliation{Jet Propulsion Laboratory, California Institute of Technology, Pasadena, CA 91109, USA}
	%\affiliation{INAF - Osservatorio Astronomico di Palermo, Piazza del Parlamento 1, I-90134 Palermo, Italy}
	
	\author[0000-0003-2215-8485]{Renyu Hu}
	\affiliation{Jet Propulsion Laboratory, California Institute of Technology, Pasadena, CA 91109, USA}
	\affiliation{Division of Geological and Planetary Sciences, California Institute of Technology, Pasadena, CA 91125, USA}
	
	\author[0000-0003-0220-0009]{Sergi R. Hildebrandt}
	\affiliation{Jet Propulsion Laboratory, California Institute of Technology, Pasadena, CA 91109, USA}
	\affiliation{Division of Physics, Mathematics and Astronomy, California Institute of Technology, Pasadena, CA 91125, USA}
	
	\begin{abstract}
		%The \textit{Starshade Rendezvous Mission} is a probe-class mission designed to fly in formation with a telescope with direct imaging capabilities. In studying exoplanets, the Starshade will block the star-light more efficiently than a classic coronagraph enabling telescope such as Roman or HabEx to reach a planet/star contrast ratio as low as $\sim$10$^{-10}$. The Starshade's operation, however, will be mainly limited by its fuel for propulsion as it will need to move between targets to align to the telescope. A careful evaluation of the scheduling for the best scientific yield is therefore a must.
		Direct imaging of widely separated exoplanets from space will obtain their reflected light spectra and measure atmospheric properties. Previous calculations have shown that a change in the orbital phase would cause a spectral signal, but whether this signal may be used to characterize the atmosphere has not been shown. We simulate starshade-enabled observations of the planet \umab, using the to-date most realistic simulator SISTER to estimate the uncertainties due to residual starlight, solar glint, and exozodiacal light. We then use the Bayesian retrieval algorithm \exorelr\ to determine the constraints on the atmospheric properties from observations using a \edit1{Roman}- or HabEx-like telescope, comparing the strategies to observe at multiple orbital phases or in multiple wavelength bands.
		With a $\sim20\%$ bandwidth in 600 -- 800 nm on a \edit1{Roman}-like telescope, the retrieval finds a degenerate scenario with a lower gas abundance and a deeper or absent cloud than the truth. Repeating the observation at a different orbital phase or at a second 20\% wavelength band in 800 -- 1000 nm, with the same integration time and thus degraded S/N, would effectively eliminate this degenerate solution. Single observation with a HabEx-like telescope would yield high-precision constraints on the gas abundances and cloud properties, without the degenerate scenario.
		These results are also generally applicable to high-contrast spectroscopy with a coronagraph with a similar wavelength coverage and S/N, and can help design the wavelength bandwidth and the observation plan of exoplanet direct imaging experiments in the future.
		%In this work, we simulated realistic reflected spectra (at different wavelength bands and at multi planetary phase angles) of a direct imaged planet observed by the system Starshade+telescope, and we used our Bayesian retrieval algorithm \exorelr\ to determine which combination of observed scenarios will provide the best scientific yield. 
		%We show that while Roman will require multiple observations to meaningfully study the atmosphere of the planet, HabEx will be able to better constrain the parameter of the retrieval with a single observation.
	\end{abstract}
	
	
	\section{Introduction} \label{sec:intro}
	
	Direct imaging of exoplanets has begun to pick up momentum as a way to characterize exoplanetary atmospheres. High-contrast observations from the ground have measured thermal emission spectra of several exoplanets \citep{Janson2010,Konopacky2013,Macintosh2015,Wagner2016,Rajan2017,Samland2017}, and the spectra have shown the presence of carbon monoxide and water in their atmospheres. These planets are young giant planets so that they emit detectable thermal emission despite being well separated from their host stars.
	
	%The direct image technique has been used for more than a decade to discover and characterize extra solar planets. The first spectrum obtained with direct imaging was recorded using the instrument NACO mounted on ESO Very Large Telescope (VLT) \citep{Janson2010}. Subsequent observations of the same target (HR 8799c) have highlighted the presence of carbon monoxide and water in its atmosphere \citep{Konopacky2013}. Using this technique, 51 Eri b \citep{Macintosh2015, Rajan2017, Samland2017} and HD 131399Ab \citep{Wagner2016} have also been studied.
	
	Spaceborne direct-imaging capabilities in the future will enable atmospheric characterization in the reflected light. The \textit{\edit1{Nancy Grace Roman Space Telescope}} \edit1{(\textit{Roman} hereafter, previously know as \textit{Wide-Field InfraRed Survey Telescope}}, \cite{Spergel2015,Akeson2019}) will be capable of collecting starlight reflected by giant exoplanets through high-contrast imaging. The \textit{Starshade Rendezvous with \edit1{Roman}} \citep{Seager2019}, an advanced mission concept, would further enable \edit1{Roman} to reach a planet-star contrast ratio as low as $\sim$10$^{-10}$. The \textit{\edit1{Habitable Exoplanet Observatory}} (HabEx, \cite{Gaudi2020}), a concept of a 4-m space telescope with a starshade, has the main objective to image \edit1{potentially Earth-like planets} and study their atmospheres. Starshade Rendezvous and HabEx, both with a starshade, would allow studying giant exoplanets orbiting their host stars at about 1 -- 10 AU with reflected light spectroscopy. 
	
	Reflected light spectra from these cold gaseous planets contain rich information on the chemical composition and cloud formation in their atmospheres \citep[e.g.,][]{Sudarsky2000, Sudarsky2003, Burrows2004, Cahoy2010, Lupu2016, MacDonald2018, Hu2019B2019ApJ...887..166H, Damiano2020,carrion2020directly}. When the wavelength coverage is limited and with moderate wavelength resolution and signal-to-noise ratio (S/N), there can be substantial degeneracy between the atmospheric abundance and the cloud pressure \citep[e.g.,][]{Lupu2016,Nayak2017,Hu2019B2019ApJ...887..166H,carrion2020directly}. In short, a deep cloud and a small abundance of an absorber (e.g., CH$_4$) may result in a similar spectrum as a shallower cloud and a higher abundance of absorber, leading to the degeneracy.
	
	A potential strategy to break the degeneracy is to observe the planets at multiple orbital phase angles. This is because viewing the same atmosphere at different phase angles would produce a wavelength-dependent difference in the spectra as the illumination and emerging angles have chromatic effects in atmospheric absorption and cloud reflection \citep{Cahoy2010, Madhusudhan2012B2012ApJ...747...25M,Nayak2017}.
	
	The reflected light spectra of Jupiter measured by telescopes from the center of the planet to the limb has led to constraints of the position and the vertical extent of the planet's upper cloud layer \citep[e.g.,][]{Sato1979}. Whether a similar measurement is as informative for exoplanets -- disk integrated but viewed at different phase angles -- is not well understood. \edit1{A previous study \citep{Nayak2017} was focused on modeling the relationship between the phase angle and other parameters such as cloud properties, methane abundance and planetary radius by considering the \edit1{Roman} coronagraph set-up. Taking into account single observations, the authors found that the knowledge of a precise phase angle does not improve the estimates for methane and cloud properties. It instead helps to improve the constraints on the planetary radius by a factor of two. In that study, the authors considered the combination of observations at different phase angles by using the intersection criterion --- a practical approximation of the exploration of parameters space without employing a Bayesian statistical search. Finally, they concluded that multiple observations at different phase angles might be beneficial for a better constraint on the physical/chemical planetary parameters.}
	
	Here, we evaluate the multi-orbital-phase observations as a strategy to refine the constraints on the atmospheric properties from reflected light spectroscopy, and compare it to multi-wavelength-band observations\edit1{ with a starshade. We focused our attention on the realistic noise model of the instruments (i.e., telescope plus starshade) and on possible degenerate scientific interpretations of the results.} Recognizing realistic operational overheads (e.g., the starshade would need to re-position itself precisely between the telescope and the star and this maneuver costs fuel), we specifically determine which of the following would produce the optimal science return when repeated observations are possible for an object of interest: \edit1{(1) repeat the observation in the same band but at a different phase, (2) repeat the observation at the same phase but in a second band, and (3) repeat the observation at the same phase and in the same band.} We use the \textit{Starshade Imaging Simulation Toolkit for Exoplanet Reconnaissance}\footnote{\url{http://sister.caltech.edu}} (\sister) to simulate data and uncertainties obtained in starshade-enabled spectroscopic observations. We then use a robust Bayesian inverse retrieval method \citep[\exorelr,][]{Damiano2020} to determine and compare the information rendered by the three strategies.

	The paper is organized as follows: Sec. \ref{sec:model} describes the simulation of the observed spectra and the multi-phase spectral retrieval method, Sec. \ref{sec:result} shows the results and compares the posterior constraints among the observation scenarios, and Sec. \ref{sec:final} discusses the observational strategy and summarizes the key findings of this study.
	
	\section{Methods} \label{sec:model}
	
	\subsection{Simulated Planet Spectra}
	
	We use \umab, a giant planet orbiting a G0V star at 2.1 AU, as the representative planet in this study. \umab\ is in the design reference mission of both Starshade Rendezvous~\citep{Seager2019} and HabEx~\citep{Gaudi2020}. Compared to Jupiter, \umab\ orbits closer to its host star and has a higher equilibrium temperature. In terms of cloud structure, this means that the upper atmosphere of the planet is expected to contain water clouds, as opposed to the ammonia clouds typical of Jupiter \citep{Sudarsky2000}. We simulated the atmospheric abundance and cloud structure of this exoplanet in our previous work \citep{Damiano2020}. %Even though, we did not explore the effects of the phase angle on the planetary albedo, our algorithm, \exorelr\, was already capable of calculating the planetary reflectivity at any angles. 
	Here we use \exorel\ \citep{Hu2019B2019ApJ...887..166H,Damiano2020} to synthesize the reflected light spectra at the orbital phases of $\pi/3$ and $\pi/2$. These phases are picked because the phase of $\pi/3$ is a good compromise between the angular separation between the planet and the star and the brightness in the reflected light (and thus the go-to phase of a single observation). The phase of $\pi/2$ is when the angular separation between the star and the planet is maximized. Should the planet be a Lambertian sphere, the brightness at $\pi/2$ is $\sim$50\% of that at $\pi/3$; but our model calculates the wavelength-dependent phase function produced by clouds and gases in the atmosphere. 
	\edit1{It is assumed in this work that the planet's atmosphere is the same at all longitudes, i.e., the clouds are not localized and the physical structure of the cloud does not change between phases.}
	%The difference in the albedo of two phase angles originates from the different part of the atmosphere probed. It is a geometrical problem which also have a chromatic effect on the albedo spectrum \citep{Cahoy2010}.
	
	% 	\begin{deluxetable}{c|c}
	% 		\tablecaption{Relevant parameters used to model the albedo spectrum of 47 UMa b. \label{tab:planet_par}}
	% 		\tablehead{
	% 			\colhead{Stellar parameter} 			  & 			\colhead{47 UMa}}
	% 		\startdata
	% 		$R_{\star}$ (R$_{\odot}$)			&			$1.24 \pm 0.04$			\tablenotemark{1}			\\
	% 		$M_{\star}$ (R$_{\odot}$)			&			$1.03 \pm 0.05$			\tablenotemark{1}			\\
	% 		$T_{eff}$ (K) 								&			$5892 \pm 70$				\tablenotemark{1}		\\
	% 		$Metallicity$ [M/H]						&			$1.5$								\tablenotemark{3}		\\
	% 		\hline
	% 		Planetary parameters 						&	  	47 UMa b									\\
	% 		\hline
	% 		$M_p$ (M$_{Jup}$)				&			$2.53 \pm 0.07$						 \tablenotemark{2}	\\
	% 		$a$ (AU) 												&			$2.1 \pm 0.02$				\tablenotemark{2}	\\
	% 		$e$															&			$0.032 \pm 0.014$	\tablenotemark{2}			\\
	% 		$Log(g)$ (Log(cm/s$^2$))					&			$3.654$	\tablenotemark{3}														\\
	% 		$T_{internal}$ (K)									&			$110$	\tablenotemark{3}																\\
	% 		$\alpha$										&			$\pi/3$ \& $\pi/2$	\tablenotemark{3}																\\
	% 		\enddata
	% 		\tablecomments{$^1$\cite{Fuhrmann1997}, $^2$\cite{Butler1996}, $^3$assumed}
	% 	\end{deluxetable}
	
	\subsection{Simulated Uncertainties} \label{sec:simulation}
	
\begin{figure} [htbp]
		\centering
		\includegraphics[scale=0.265]{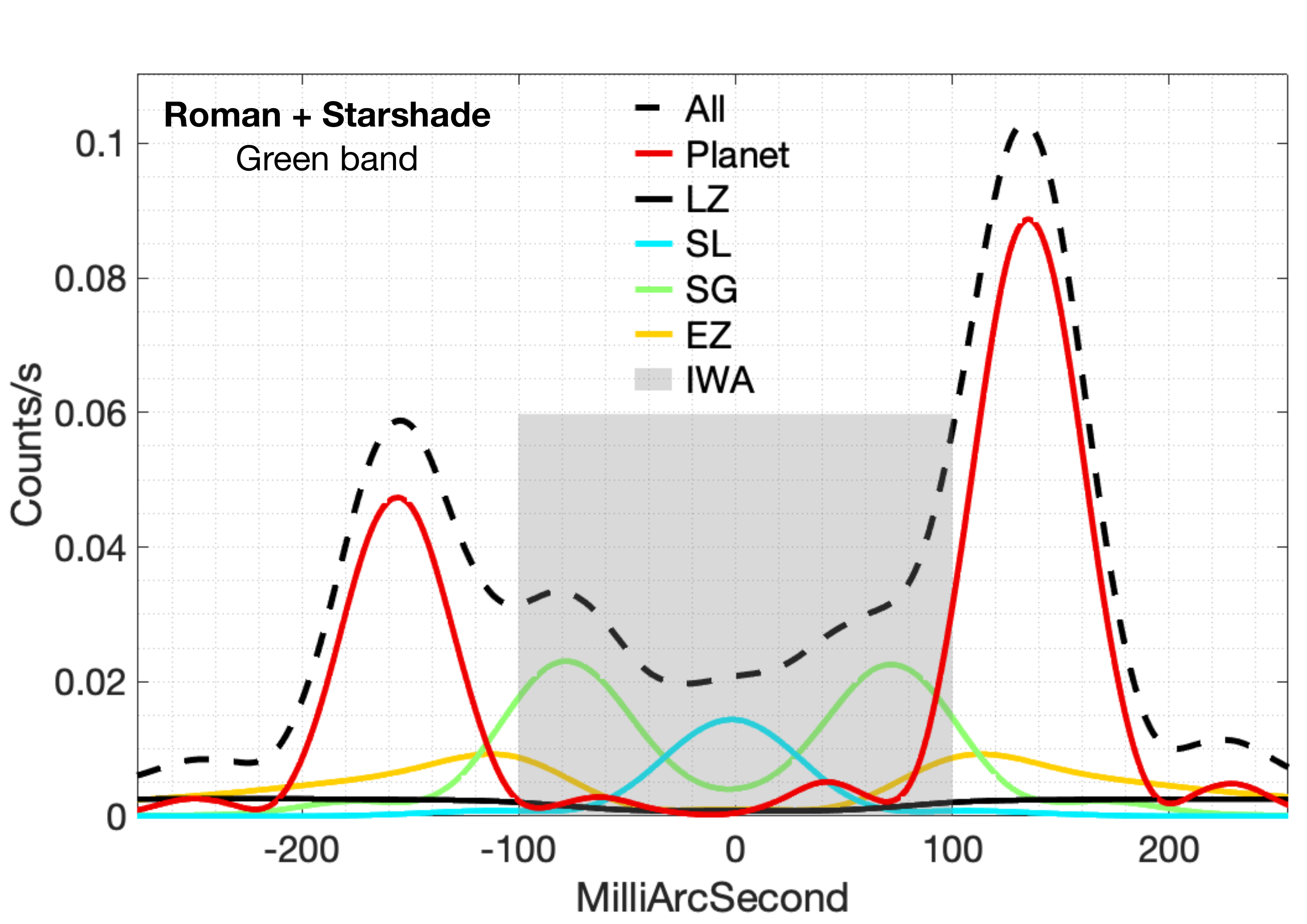}
		\includegraphics[scale=0.265]{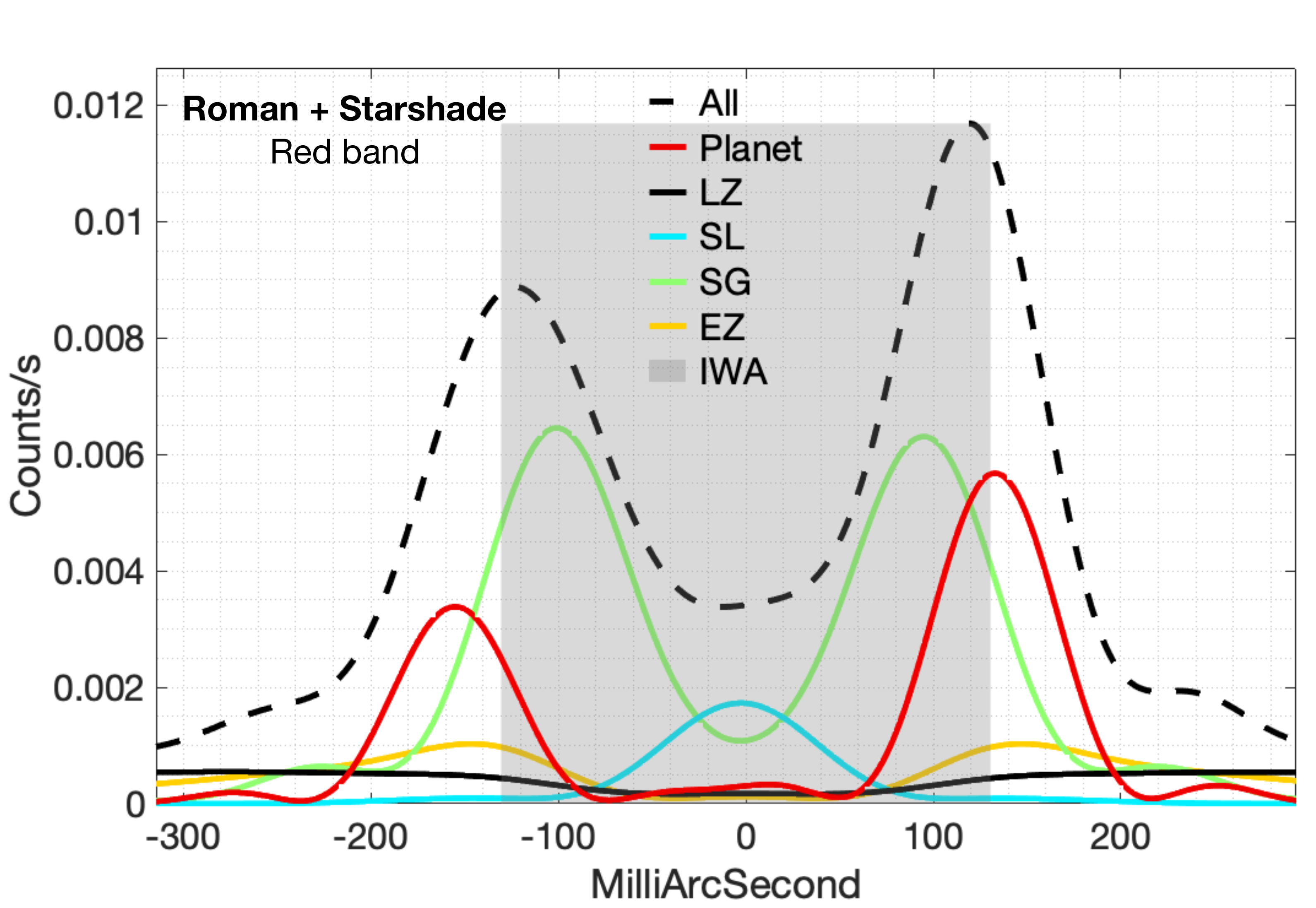}
		\caption{Contribution of each astrophysical and background component to the Starshade Rendezvous simulation performed by SISTER. The unit on the vertical axis is the rate of photons arriving at the instrument's detector taking into account all the optical and instrumental effects, except noise, and has been integrated over the bandwidth. {\bf All} is the sum of all individual contributions. {\bf Planet} refers to the rate coming from the planet only. We plot on the left-hand side the planet when its orbital phase is $\pi/2$, and on the right-hand side when it is $\pi/3$. {\bf LZ} refers to the contribution from the local zodiacal light at 22.5 V-mag/arcsec${}^2$. {\bf SL} is the contribution of the residual starlight from starshade imperfections. {\bf SG} refers to the contribution from the solar glint for a representative angle of the Sun with respect to the target star. {\bf EZ} is the contribution from the exozodiacal emission, which was chosen to be 5 times the \edit1{dust density} of the Solar System. Finally, {\bf IWA} refers to the geometric Inner Working Angle and is the apparent size of the starshade as seen by the telescope. Diffraction effects are relevant close and inside the IWA. Top: 0.61--0.75 $\mu$m. The planet-star flux ratio averaged over the bandwidth is 1.9$\times10^{-8}$ when the orbital phase is $\pi/3$ and 1.0$\times10^{-8}$ when it is $\pi/2$. The IWA is 100.2 mas. Bottom: 0.82--1.0 $\mu$m. The average planet-star flux ratio is 7.6$\times10^{-9}$ when the orbital phase is $\pi/3$ and 4.2$\times10^{-9}$ when it is $\pi/2$. The IWA is 130.4 mas. Notice that the photon rate is lower than in the 0.61--0.75 $\mu$m band.} 
		\label{fig:sister_lines_no_cumulative}
	\end{figure}

	A novel aspect of this study is the inclusion of realistic uncertainties expected from observations using a starshade.
	We use \sister\ (Hildebrandt et al., in prep) to simulate the photons' trajectory in the starshade-telescope 
	system as well as detector effects in the
	telescope. Exposure time calculators \cite[e.g.,][]{Robinson2016} have been developed to estimate the uncertainties of a starshade 
	or coronagraph observation. However, they did not take into account the full 2-dimensional nature of the 
	astrophysical scene nor the spatial variation of the Point Spread Function (PSF) 
	due to the optical diffraction from the starshade. SISTER contemplates these aspects. 
	Most details of SISTER can be found in its 
	handbook\footnote{\url{http://sister.caltech.edu}}.  
	Our simulated astrophysical scene includes \umab, residual starlight, exozodiacal light (five times dustier than the Solar System, \cite{Ertel2020}), solar glint (i.e., scattered and reflected light from the Sun by the starshade, \cite{Hilgemann2019}), and local zodiacal light.
	Specifically, we consider petals of the starshade to have imperfection compatible with current lab work, which results in a greater residual starlight than an ideal starshade. We then place the planet as close as possible to the maximum intensity of the solar glint compatible with its orbital parameters.	As such, we generate a scenario that serves as an upper limit for the integration time compared to other more 
	favorable scenarios of the same system. Figure~\ref{fig:sister_lines_no_cumulative} shows the (noiseless) contribution of each component arriving at the detector.
	
	On the telescope side, the optical throughput and the quantum efficiency (QE) of the detector follow the expected performance
	of \edit1{Roman} and HabEx, respectively. The detector models an Electron Multiplying Charge Couple Device (EMCCD). The EMCCD gain factor is set to 1,000, making the read out noise effectively zero.
	We choose noise parameters corresponding to those adopted by \edit1{Roman} and HabEx, respectively, including clock induced charge and dark current~\citep{Nemati2017}.
	
	When computing the uncertainties in the planetary counts, we assume that residual starlight, solar glint, exozodiacal light, 
	and local zodiacal light can be subtracted out with some post-processing technique
	to the shot-noise limit, i.e., leaving no systematic 
	bias in the measurements.
	This assumption increases the shot noise of these components 
	by $\sim \sqrt{2}$ as a consequence of subtracting a noisy template from noisy data.
	The final error bars for each spectral bin are then derived from 1,000 independent SISTER simulations
	and used to generate independent Gaussian noise realizations, which are finally added to the input spectral data points.

	We now turn to the specific scenarios that correspond to \edit1{Roman} and HabEx.
	
	\subsubsection{Starshade Rendezvous with Roman}
	
	We consider a \edit1{Roman}-like telescope with an Integral Field Spectrograph (IFS) that operates in 3 separate wavelength bands 
	in 0.4 -- 1.0 $\mu$m with a 20\% bandwidth. We specifically assume ``blue'' (0.45--0.55 $\mu$m), ``green'' (0.61--0.75 $\mu$m), and ``red'' (0.82--1.0 $\mu$m) bands,
	all with resolving power $R=50$. 
	These bands do not match exactly with the \edit1{Roman} Coronagraph's baseline design (V. Bailey, private comm.) or 
	the design reference mission of the Starshade Rendezvous concept \citep{Seager2019}, as these designs have been evolving. 
	In general, however, the blue band would cover the expected brightest section of the albedo spectrum and so is 
	the best for planet detection, the green band would cover a strong absorption band of CH$_4$ and so is the primary band 
	for atmospheric characterization, and the red band would provide additional information on CH$_4$ and cover an absorption 
	band of H$_2$O \citep[e.g.,][]{Lupu2016,MacDonald2018}. With a 20\% bandwidth, multiple observations would be necessary to obtain a spectrum that spans more than one band.
	
	We determine the integration time by requiring an average S/N of 20 per spectral bin of the blue band
	when the orbital phase is $\pi/3$. The integration time is $\sim3.5$ hours. We use this integration time in the other two bands, and in the observations 
	with an orbital phase of $\pi/2$. 
	We obtain an average S/N of 18.0 per spectral bin in the green band and 2.3 in the red band 
	when the orbital phase is $\pi/3$. 
	The S/N decreases to 11.5 and 1.5 when the orbital phase 
	is $\pi/2$ (see Figure~\ref{fig:wfirst}). The ratio of S/N between the 60-degree observation and the 90-degree observation is $\sim1.55$ while the ratio of planetary flux is $\sim2$. This is because the planetary flux contributes substantially to the shot noise in these observations, and the background fluxes from solar glint and exozodiacal dust decrease for wider angular separations from $\pi/3$ to $\pi/2$ (Figure~\ref{fig:sister_lines_no_cumulative}).
	
	\edit2{The integration times estimated here do not include the expected difference in detector QE, dark current and clock induced charge between \textit{beginning-of-life} (BOL) and \textit{end-of-life} (EOL). Roman is expected to be launched before the Starshade Rendezvous. Therefore, the QE, the dark current and the clock induced charge will degrade by the time the Starshade is coupled with Roman. If we take into account these factors, the calculated integration time would be $\sim8.7$ hours, compared with $\sim3.5$ hours. These achromatic factors do not introduce substantial differences in the S/N ratio between different wavelength bands and phase angle observations.}
	
	\subsubsection{HabEx}
	
	The HabEx concept \citep{Gaudi2020} entails a larger telescope than \edit1{Roman} that would be able to obtain 
	spectra in a broad wavelength range between $0.45-1.0$ $\mu$m with a resolving power of $R=140$ with a single observation. 
	The broader bandwidth and higher spectral resolution may provide improved constraints on atmospheric properties. The HabEx concept also entails a larger starshade than \edit1{Roman} Rendezvous, and it would be placed further away from the telescope. This results in a decrease of the solar glint background, and this effect is included in SISTER.
	Here we evaluate whether multi-phase observations would further improve the constraints from single observation. 
	The HabEx design includes an additional wavelength band at $1.0-1.8$ $\mu$m; we do not include this band in the 
	analysis because it is primarily for characterizing small planets \citep{Gaudi2020}.
	
	We simulate the uncertainties in HabEx simulations in a similar way as with \edit1{Roman}. 
	To compare the two cases, we adjust the integration 
	time to obtain an average S/N of 20 per spectral bin in the 0.45--0.55 $\mu$m interval when the orbital phase is $\pi/3$. 
	The integration time is $\sim1.8$ hours. As in the case of \edit1{Roman}, we use this integration time 
	when the orbital phase is $\pi/2$. The average S/N per spectral bin in the 0.45--0.55 $\mu$m interval 
	is 15.0 when the orbital phase is $\pi/2$ (see Figure~\ref{fig:habex}).
	
	\edit2{We note that despite the HabEx design is expected to have higher spectral resolution, larger primary mirror diameter and better QE above 750 nm, the integration time estimated here is only a factor of $\sim2$ less than the Roman scenario because we based our calculations at 500 nm where the QE of the two instruments is similar. We then used the same integration time to other wavelengths to focus our attention on the S/N ratio between different wavelengths. The benefits of HabEx are appreciated at longer wavelengths where the differences between the two instruments are substantial (see Sec. \ref{sec:result} and Fig. \ref{fig:habex}).}
	
	\subsection{Multi-Phase and Multi-Band Retrieval} \label{sec:retrieval}
	
	We augment the \exorelr\ retrieval framework \citep{Damiano2020} with the ability to retrieve multiple spectra taken at different phase angles or in different wavelength bands. 
	%In this work, we performed statistical retrieval analysis of single spectra, in this case we refer to the set-up described in \cite{Damiano2020}, and multiple spectra taken at different phase angles or in different wavelength bands. Since the multi-phase spectra contain different spectral information, we can not combine them by taking the weighted average mean. 
	To simultaneously fit the spectra within the same Bayesian instance, we calculate the likelihood of each set of data and take the product. Specifically, when a set of free parameters is chosen to be evaluated, we calculate the spectrum for the first phase angle and we compare the model with the data to obtain the likelihood $\mathcal{L}_1$. With the same set of parameters we calculate the model referred to the second phase angle and we compare it to the data to calculate the likelihood $\mathcal{L}_2$. The product, $\mathcal{L}_1 \times \mathcal{L}_2$ will give the total likelihood of the free parameters set, as
	\begin{equation}
	log(\mathcal{L}) = log(\mathcal{L}_1 \times \mathcal{L}_2) = log(\mathcal{L}_1) + log(\mathcal{L}_2).
	\end{equation} 
	
	Another update is that we now retrieve from the planet-star contrast ratio, rather than the albedo in \cite{Damiano2020}, because the uncertainties simulated are expressed in the planet-star contrast ratio. This brings in the planetary radius as a major factor. We now include the surface gravity as an additional free parameter in the retrieval, and the planetary radius is derived from the planetary mass and the retrieved surface gravity.
	
	\section{Results} \label{sec:result}
	
	\begin{deluxetable*}{c|c|cccc}
		\tablecaption{Results of the retrievals for each of the observational scenarios of Starshade Rendezvous with Roman considered for \umab. $\alpha$ is the phase angle. The reported errorbars are relative to the 95\% confidence interval. \label{tab:wfirst_res}}
		\tablehead{Parameters & Truths & \multicolumn{4}{c}{Starshade Rendezvous with Roman}}
		\startdata
		$ $ & & Green band & Green band & Green \& red bands & Green band \\
		$ $ & & $\alpha=\pi/3$ only & $\alpha=\pi/3$ \& $\pi/2$ & $\alpha=\pi/3$ & $\alpha=\pi/3$ twice \\
		\hline
		$Log(H_2O)$	& $-1.50$ & $-1.18^{+1.07}_{-10.20}$ & $-0.90^{+0.79}_{-1.60}$ & $-0.98^{+0.84}_{-1.36}$ & $-0.98^{+0.87}_{-1.21}$ \\
		$Log(NH_3)$	& $-2.36$ & $-7.46^{+6.18}_{-4.16}$ & $-6.29^{+5.65}_{-5.15}$ & $-4.50^{+3.55}_{-6.64}$ & $-6.90^{+5.49}_{-4.72}$ \\
		$Log(CH_4)$	& $-1.80$ & $-1.45^{+1.13}_{-2.61}$ & $-1.36^{+0.93}_{-1.61}$ & $-1.43^{+1.02}_{-1.67}$ & $-1.52^{+1.17}_{-1.12}$ \\
		$Log(P_{H_2O}[{\rm Pa}])$ & $3.84$ & $2.70^{+4.62}_{-2.43}$ & $3.05^{+1.78}_{-2.13}$ & $3.29^{+1.93}_{-2.15}$ & $3.53^{+1.42}_{-2.85}$ \\
		$Log(D_{H_2O}[{\rm Pa}])$	& $4.97$ & $4.58^{+3.31}_{-3.66}$ & $4.73^{+0.96}_{-0.49}$ & $4.72^{+1.04}_{-0.49}$ & $4.80^{+0.76}_{-0.64}$ \\
		$Log(CR_{H_2O})$ & $-5.04$ & $-7.93^{+6.59}_{-3.70}$ & $-7.04^{+4.64}_{-4.33}$ & $-7.36^{+4.98}_{-4.35}$ & $-8.80^{+5.98}_{-2.96}$ \\
		$Log(g [{\rm cm/s^2}])$ & $3.65$ &  $3.66^{+0.05}_{-0.04}$ & $3.68^{+0.03}_{-0.03}$ & $3.68^{+0.03}_{-0.04}$ & $3.66^{+0.03}_{-0.06}$ \\
		\enddata
	\end{deluxetable*}
	
	\begin{deluxetable*}{c|c|ccc}
		\tablecaption{Results of the retrievals for each of the observational scenarios of HabEx considered for \umab. $\alpha$ is the phase angle. The reported errorbars are relative to the 95\% confidence interval.\label{tab:habex_res}}
		\tablehead{Parameters & Truths & \multicolumn{3}{c}{HabEx}}
		\startdata
		$ $ & & $\alpha=\pi/3$ & $\alpha=\pi/3$ \& & $\alpha=\pi/3$ \\
		$ $ & & only & $\alpha=\pi/2$ & twice \\
		\hline
		$Log(H_2O)$	& $-1.50$ & $-1.60^{+0.58}_{-0.31}$ & $-1.32^{+0.78}_{-0.40}$ & $-1.30^{+0.71}_{-0.37}$ \\
		$Log(NH_3)$	& $-2.36$ & $-2.78^{+0.98}_{-8.51}$ & $-2.11^{+0.91}_{-6.43}$ & $-2.51^{+0.82}_{-8.36}$ \\
		$Log(CH_4)$	& $-1.80$ & $-1.76^{+0.36}_{-0.36}$ & $-1.59^{+0.61}_{-0.33}$ & $-1.62^{+0.51}_{-0.19}$ \\
		$Log(P_{H_2O}[{\rm Pa}])$ & $3.84$ & $2.52^{+1.50}_{-1.95}$ & $2.22^{+1.48}_{-1.95}$ & $2.44^{+1.34}_{-1.74}$ \\
		$Log(D_{H_2O}[{\rm Pa}])$	& $4.97$ & $5.00^{+0.24}_{-0.16}$ & $4.96^{+0.16}_{-0.24}$ & $4.96^{+0.07}_{-0.16}$ \\
		$Log(CR_{H_2O})$ & $-5.04$ & $-8.75^{+4.94}_{-3.04}$ & $-9.21^{+4.77}_{-2.65}$ & $-9.18^{+4.88}_{-2.69}$ \\
		$Log(g [{\rm cm/s^2}])$ & $3.65$ & $3.65^{+0.02}_{-0.01}$ & $3.66^{+0.02}_{-0.01}$ & $3.66^{+0.02}_{-0.01}$ \\
		\enddata
	\end{deluxetable*}
	
\begin{figure*}[!h]
		\centering
		\includegraphics[angle=90, scale=0.575]{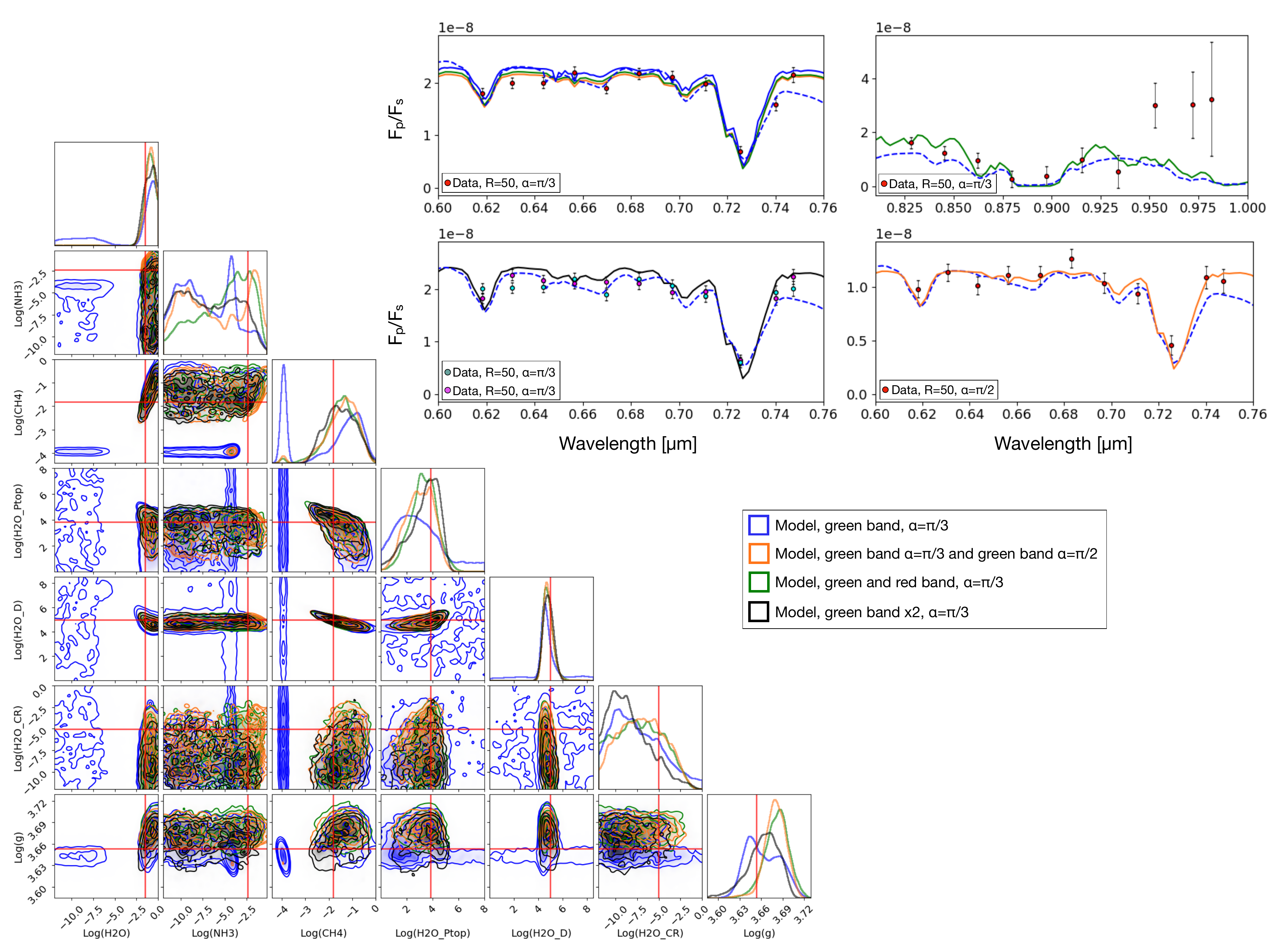}
		\caption{\label{fig:wfirst}Retrieval analysis of the four observational scenarios of Starshade Rendezvous with \edit1{Roman}. The four panels on the top show the best-fit models and the simulated data. The dashed blue line in the first panel shows the degenerate solution obtained with a single observation in the green band and at the orbital phase of $\pi/3$; the other three panels show how this solution may be inconsistent with additional observations. The corner plot on the bottom shows the marginalized posterior distribution for each of the parameters, using the same color scheme to denote the four observational scenarios simulated. The degenerate solution is eliminated with a second observation at the same phase, at a wider-separated phase of $\pi/2$, or in the red band.}
	\end{figure*}

\begin{figure*}[!h]
		\centering
		\includegraphics[angle=90, scale=0.575]{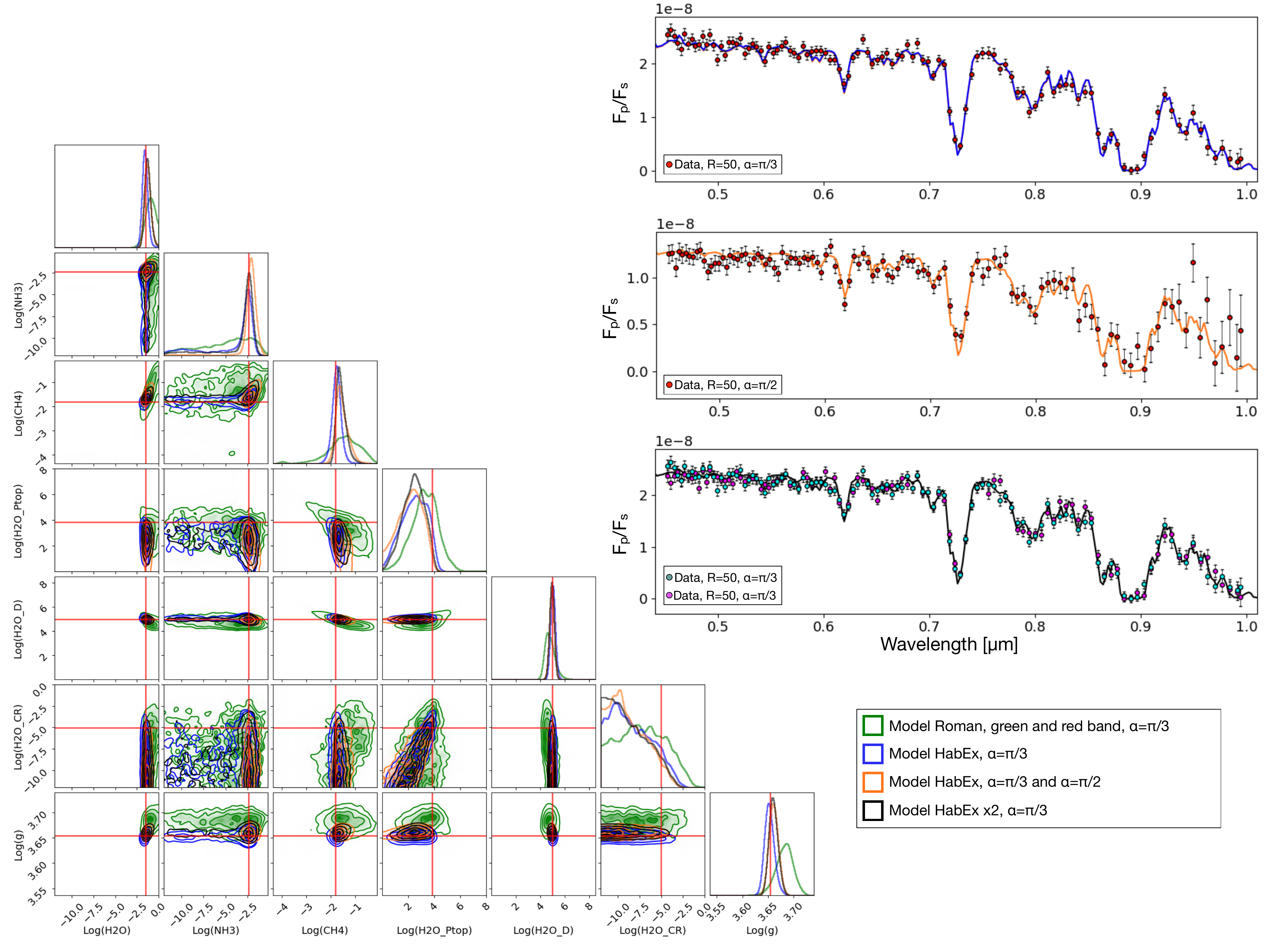}
		\caption{\label{fig:habex}Retrieval analysis of the three observational scenarios of HabEx. The three panels on the top show the best-fit models and the simulated data. The corner plot on the bottom shows the marginalized posterior distribution for each of the parameters, using the same color scheme to denote the three observational scenarios simulated. The green model in the corner plot is the same as the green model in Figure~\ref{fig:wfirst}, and it is shown for comparison. A single observation with HabEx would be sufficient to constrain the atmospheric properties.}
	\end{figure*}

	The constraints on the atmospheric properties retrieved from the simulated observations with Starshade Rendezvous with \edit1{Roman} and HabEx are summarized in Tables \ref{tab:wfirst_res} and \ref{tab:habex_res} and shown in Figures \ref{fig:wfirst} and \ref{fig:habex}. Definitions of these parameters and their impact on the albedo spectrum are described in detail in \cite{Damiano2020}. Briefly, $Log(H_2O)$, $Log(NH_3)$, and $Log(CH_4)$ are the mixing ratios of the trace gases; since we include condensation of H$_2$O, the mixing ratio of H$_2$O is the mixing ratio below the cloud. $Log(P_{H_2O})$ and $Log(D_{H_2O})$ are the top pressure and depth of the H$_2$O cloud. The cloud depth is the difference in pressure between the bottom and the top of the cloud. $Log(CR_{H_2O})$ is the ratio between the mixing ratio of H$_2$O above the cloud and that below the cloud -- the difference is condensed out to form the cloud.
	
\begin{figure*}[]
		\plotone{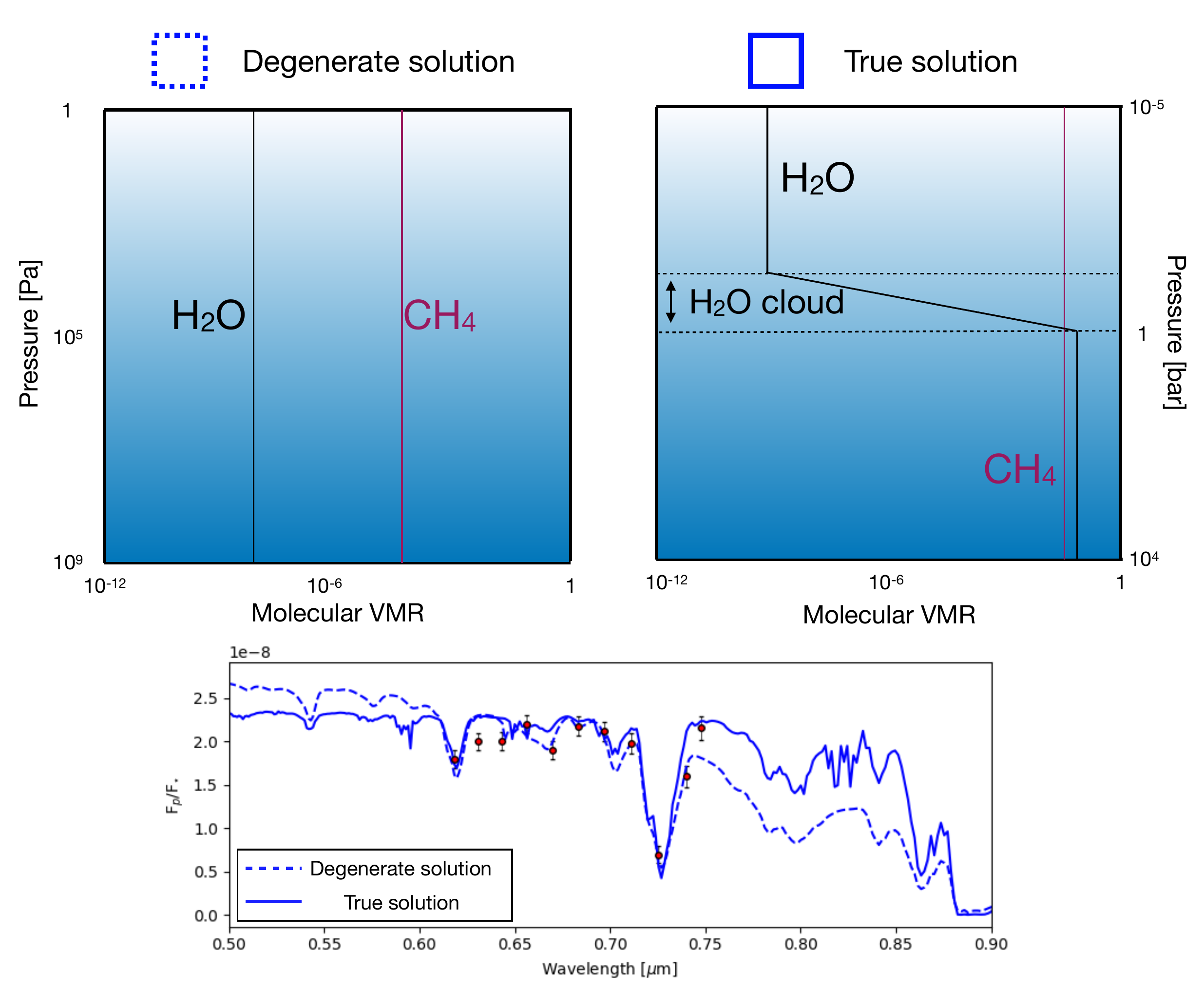}
		\caption{\edit1{\textbf{Top panels:} atmospheric composition of the degenerate solution (on the left) and the true solution (on the right). The degenerate solution represents a clear sky scenario with low abundance of water and methane. The true solution, on the other hand, comprises water cloud at around 1 bar. Water and methane concentrations are higher than the degenerate solution. \textbf{Bottom panel:} the spectra of the two scenarios (dashed line for the degenerate scenario and solid line for the true one) are represented together with the simulated data (green band, $\alpha=\pi/3$). The difference between the two spectra is large at shorter and longer wavelengths.} \label{fig:degenerate}}
	\end{figure*}

	\textbf{\edit1{Roman}.} The most plausible starting point to characterize the atmosphere of \umab\ in reflected light is a single observation in the green band. Our retrieval analysis in this scenario finds a solution close to the truth, as well as a degenerate solution (dashed blue lines in Figure~\ref{fig:degenerate} and \ref{fig:wfirst}). While the truth has a mixing ratio of CH$_4$ of $\sim10^{-2}$ and a H$_2$O cloud at $0.1\sim1$ bar, the degenerate solution has a mixing ratio of CH$_4$ of $\sim10^{-4}$ and corresponding mixing ratio of H$_2$O $<10^{-6}$ (Figure~\ref{fig:wfirst}). With such a low abundance of water, there is not enough mass to form a reflective cloud, and so the degenerate solution is insensitive to the cloud description parameters ($Log(P_{H_2O})$, $Log(D_{H_2O})$, and $Log(CR_{H_2O})$). The degenerate solution is essentially a cloud-free atmosphere with the mixing ratio of CH$_4$ two-order-of-magnitude lower than the truth.
	
	%1) The first scenario we considered is the single observation of the green band at phase-angle equal to $\pi/3$. This will represent the initial observation of a possible campaign. We performed the retrieval analysis on the data points and we found a degenerate solution (see Fig. \ref{fig:wfirst} blue model). The Bayesian sampler found a solution close to the truth values used to generate the spectrum in the first place, but it also found a second solution that also fits the data. The degenerate solution represents an atmosphere with low concentration of methane and water (weak absorption) with practically no clouds (Fig. \ref{fig:wfirst} dashed blue model). 
	
	Adding a second observation at a different orbital phase ($\alpha=\pi/2$ in this study) with the same integration time would drastically reduce the likelihood of the degenerate solution. The S/N of the second observation is poorer, because the baseline of the spectrum is lower at the phase of $\alpha=\pi/2$ than at the phase of $\alpha=\pi/3$. The simultaneous retrieval of the observations at both phases would find the true result (see the orange model and distributions shown in Figure~\ref{fig:wfirst}). The improvement in the constraints comes from two factors. First, adding a second observation inevitably increase the S/N of the overall observation, which helps eliminate the degenerate solution. Second, the true and the degenerate scenarios have different achromatic phase functions, and the biggest difference occurs at the main absorption bands (Figure~\ref{fig:wfirst}). The difference is on the order of 5\%, so that the S/N of 11.5 per spectral bin from the assumed observational scenario could partly leverage this diagnostic power.
	
	% 2) For the second scenario, we added a second observation of the same wavelength band but at different phase-angle ($\alpha=\pi/2$). The simultaneous retrieval of both the observations resulted again in a degenerate result. However, the degenerate solution this time is less significant (see Fig. \ref{fig:wfirst} orange model). 
	
	Moreover, adding an observation in the red band with the same integration time would also drastically reduce the significance of the degenerate solution almost to zero. Here we assume that the second observation takes place immediately after the first one, and they essentially correspond to the same orbital phase. The red-band observation has an average S/N of only $\sim2$, due to decreasing starlight intensity, lower planetary albedo, stronger solar glint, and degrading detector quantum efficiency from the green band to the red band. Despite the poor S/N of the red-band spectrum, we find it highly informative when combined with the green-band spectrum. The retrieval of the observations at both bands does not show a degenerate solution (see the green model and distributions shown in Figure~\ref{fig:wfirst}). This is because the degenerate scenario has a much lower albedo than the truth in the red band, and the difference is well on the order of 30\% in 0.75 -- 0.85 $\mu$m.
	
	%By using the same phase-angle, we intended to simulate an observation immediately subsequent to the first one instead of waiting for the planet to move in its orbit. We performed a simultaneous retrieval on the two wavelength band data. The result does not show any degenerate solutions this time. In Fig. \ref{fig:wfirst} the dashed blue model (i.e. the degenerate solution in scenarios 1) and 2)) does not encompass well the data of the red band, and for this reason in this scenario it is not a solution. 
	
	The last scenario considered is to repeat the first observation at the same wavelength band and at the same orbital phase, i.e., a second realization. The combined S/N is $\sim25$ per spectral bin, and to our surprise, this strategy would also eliminate the degenerate solution. Most of the diagnosing power comes from the datum at 0.75 $\mu$m, where the truth and the degenerate scenario differ the most. 
	
	%scenario represents two observations of the same wavelength band at the same phase-angle to simulate two subsequent observation of exact duration in the same wavelength range. Also in this case we did not find any degenerate solution. 
	
	%In the first three scenarios we noticed a small bias on the retrieved value of CH$_4$, however, the true value is still encompassed by the probability density function (PDF).
	As long as the degenerate solution is eliminated, the reflected light spectral retrieval can constrain not only the mixing ratio of CH$_4$ in the atmosphere, but also the mixing ratio of H$_2$O below the cloud (as it is the feedstock for cloud), and the pressure level of the cloud. There is a moderate correlation between the mixing ratio of CH$_4$ and the pressure level of the cloud (Figure~\ref{fig:wfirst}). Inferring the mixing ratio of H$_2$O below the cloud is possible because our retrieval method preserves the causal relationship between the condensation of water vapor and the formation of a water cloud \citep{Damiano2020}. The surface gravity, and thus the radius of the planet is constrained in all scenarios.
	
	%the amount of water below the clouds, the amount of methane in the atmosphere, the position of the clouds and the gravity of the planet in the third and forth scenarios. 
	
	\textbf{HabEx.} A single observation of HabEx with a S/N of $\sim20$ would pinpoint the atmospheric properties (see the blue model and distributions shown in Figure~\ref{fig:habex}). Specifically, the mixing ratios of CH$_4$ and H$_2$O would be well constrained to the precision of $\sim1/3$ dec, and that of the cloud pressure would be better constrained than $\sim1/5$ dec. There is no correlation between the mixing ratio of CH$_4$ and the cloud pressure in this case. Notably, the spectrum yields meaningful constraints on the mixing ratio of NH$_3$, mostly from the small absorption feature at $\sim0.63$ $\mu$m. These improvements come from not only the broader wavelength coverage and the higher spectral resolution than \edit1{Roman}, but also the larger and further starshade (for lower solar glint) and the larger telescope aperture (for lower exozodiacal light). Adding another observation with at the same or a different orbital phase would slightly reduce uncertainty in the retrieved mixing ratios of CH$_4$, while the overall gain in the constraints of the parameters is minimal.

	\section{Discussion \& Conclusion} \label{sec:final}
	
	%In our previous work \citep{Damiano2020}, we did not focus about the effect of the data noise on \exorelr\ retrieval performances.
	%In this work we presented real-life case scenarios of possible observations done by the Roman and HabEx telescopes and their respective Starshades. We simulated realistic observations by using the combination of \exorelr\ and SISTER. Since the Starshade mission will need a strict observational scheduling due to its limited propellant, the goal of this work is to explore the possible strategies that increase the science return. To perform this experiment we have chosen the cold jovian planet \umab as it is one of the candidate planet to be observed by both telescope.
	
	With the case studies presented in Section~\ref{sec:result}, we show that with a limited wavelength band like the green band on \edit1{Roman}, a single observation of a planet's reflected light spectrum at the favorable orbital phase (e.g., $\alpha=\pi/3$) and a high S/N (e.g., $\sim$20) can result in degenerate solutions. The degeneracy is between the atmospheric scenario of a higher cloud and greater mixing ratio of CH$_4$ versus the scenario of a lower or absent cloud and smaller mixing ratio of CH$_4$. This type of degeneracy has been discussed in \cite{Hu2014B2014arXiv1412.7582H,Hu2019B2019ApJ...887..166H} and it is shown here with realistic noise estimates and rigorous retrieval.
	
	Expanding the wavelength coverage to longer wavelengths (i.e., the red band), doubling the integration time, or repeating the observation at a different orbital phase with a similar integration time would be able to eliminate this degeneracy. As shown in Section~\ref{sec:result}, these three strategies result in similar-quality constraints on the atmospheric parameters.
	
	Particularly, the spectra that are degenerate in the green band diverge at longer wavelengths: the low or absent cloud scenario would have a much lower albedo than the high cloud scenario. The divergence is often so significant that even a low S/N observation at the long wavelengths can effectively eliminate the degenerate solution.
	To date, it is often assumed that the red-band observation with \edit1{Roman} would be severely affected by the degrading quantum efficiency of the detector in this band \citep[e.g.,][]{Seager2019}. Here we show that even with the poor quantum efficiency and other complications, a red-band observation with the same integration time as the green band can be highly informative. This finding also implies that the constraints on the atmospheric properties would be drastically improved at no additional integration time, if the wavelength bandwidth could be enlarged to cover both the green band and the red band with a single observation.
	
	If broadening the wavelength coverage is not possible, eliminating the degenerate solution would require an S/N per spectral bin of at least $\sim25$, which may be achieved by integrating longer in a single observation or revisiting the same planet at a different orbital phase. At an S/N of $\sim18$ per visit, the difference between revisiting and integrating longer is minimal, and thus integrating longer may be the preferred strategy to avoid re-targeting the starshade. As shown in Fig. \ref{fig:wfirst}, there is indeed a difference in the spectral features between the two phases, consistent with \cite{Cahoy2010}. To capture this subtle difference, however, will require a high S/N ($>20$) on the observations at both orbital phases and thus a longer integration time for the observation at the less-than-optimal phase. The analysis here highlights that, when only one observation in a limited wavelength bandwidth is expected, it is important to schedule that observation at the phase that maximizes the S/N of the spectrum and with sufficient duration.
	
	\edit1{In this work, we have assumed the phase angle of the observations and the mass of the planet to be known. In this context, the planetary gravity, that we considered as a retrievable parameter, is effectively derived from the radius of the planet. This assumption for the characterization of giant planets is acceptable as observational campaigns and radial velocity observations may provide a good constraint on the aforementioned parameters. However, if the phase angle and/or the mass of the planet are poorly known, the planetary radius would become a free parameter detached from the gravity, and one would expect correlations between the planetary radius and the phase angle \citep{Nayak2017}.}
	
	\edit1{In this project, when considering multi-phase-observations and multi-band-observations, we focused our attention on the noise ratio between the scenarios by using the same integration time for all the observations. This means that in this work we combined observations very different in terms of S/N. In spite of this fact, we obtained improvements on the constrain of the free parameters when multiple observations have been combined. In line with previous work \citep{Nayak2017}, the improvements generally do not exceed the order of magnitude, with the most significant improvements in the methane abundance and in the cloud top pressure constraints.}

	With a wide wavelength coverage in the visible and near-infrared, a HabEx-like telescope would provide precise constraints on the atmospheric gas abundances and cloud properties in widely separated giant exoplanets with one observation of the reflected light, without the degeneracy between the gas abundance and the cloud pressure as discussed above. Not knowing the planetary radius does not substantially affect this capability. To our knowledge, this is the first time such capability is {\it demonstrated} with realistic uncertainty estimates and Bayesian retrievals. Also, as we have shown in \cite{Damiano2020}, the abundance of the cloud-forming gas (e.g., H$_2$O in \umab) below the cloud can be constrained from the reflected light spectra.
	
	\edit1{Finally, \exorelr\ has been designed to be applicable to the population of cold gaseous planets orbiting at the semi major axis of \textgreater1 and \textless10 AU from solar-like stars. These planets may be described with a single or double layers of clouds, made of water and/or ammonia ices depending on temperature and optical depth. 47 Uma b, the target considered in this work, is one of the planets in this category. The degenerate solutions and the strategies of mitigation presented here are thus broadly applicable to the planets in the aforementioned category.}
	
	%The discussion is different for HabEx. The capabilities of the telescope are better than Roman given its higher spectral resolution and broader wavelength coverage. We found that a single observation is enough to constrain most of the parameters ($P_{top, H_2O}$ and $CR_{H_2O}$ do not contribute much to the planetary albedo as defined and shown in \cite{Damiano2020}). Multiple observations (subsequent or at significant different phase-angles) will not result in a meaningful improvement.
	
	In conclusion, we present simulations of starshade-enabled reflected light spectroscopy of widely separated giant exoplanets and use Bayesian retrievals to assess strategies to improve the constraints on the atmospheric properties. We show that a degenerate scenario with a lower gas abundance and a deeper or absent cloud than the truth can be retrieved when the observation is limited to a $\sim20\%$ band in 600 -- 800 nm and has an S/N of $\sim18$. Widening the wavelength coverage to include 800 -- 1000 nm is the most effective way to eliminate the degenerate solution. Doubling the integration time or repeating the observation at a different, less-than-optimal orbital phase would also eliminate the degenerate solution. \edit1{However, observation overheads must be taken into consideration as the mission resources are limited. Doubling the integration time likely comes with very little overhead, but costs twice as much station-keeping fuel for the starshade. Moreover, the starshade would need to be moved closer to the telescope to repeat the observation in the red band. The multi-phase option requires doubling overheads and the fuel associated with slewing the starshade back to the same target.} Although these results are based on starshade simulations, they are generally applicable to high-contrast imaging and spectroscopy with a coronagraph with similar wavelength coverage, spectral resolution, and S/N. The work presented here will be important to design and define the wavelength bandwidth and the observation plan of exoplanet direct imaging experiments in the future.
	
	%The work we presented here is important to design and define the Starshade schedule. Similar evaluation should be conducted in all the possible targets to determine \textit{a priori} the combination of the observations that result in the best scientific yield. Our result on Roman and HabEx highlights that a target can be studied without moving the Starshade for multiple phase-angle observations saving propellant and optimize in this way the mission lifetime. 
	
	\section*{Acknowledgments}
	The authors thank Dr. Gra\c{c}a M. Rocha for helpful discussions in the preparation of this manuscript, \edit2{Dr. Stefan Martin for providing the optical throughput and the detector quantum efficiency used by \edit1{Roman} Coronagraph and HabEx, and Dr. Andrew Romero-Wolf for providing Roman's QE value at EOL. Finally, we also thank the anonymous referee for the valuable comments provided which greatly helped to improve the manuscript.} This work was supported in part by the NASA WFIRST Science Investigation Teams grant \#NNN16D016T. This research was carried out at the Jet Propulsion Laboratory, California Institute of Technology, under a contract with the National Aeronautics and Space Administration. 

	{	\small
		\bibliographystyle{apj}
		\bibliography{bib.bib,bib2.bib}
	}

\end{document}